\RequirePackage{fix-cm}
\documentclass[smallextended]{svjour3}       
\smartqed  
\usepackage{graphicx}
\usepackage{color}
\usepackage{listings}
\usepackage{booktabs}
\usepackage{array}

\usepackage{subfigure}

\usepackage{cite}
\usepackage{url}
\usepackage{fancyhdr}

\usepackage{soul}

\usepackage{float}
\usepackage{hyperref}
\usepackage{lineno}
\usepackage{listings}
\usepackage{mdwmath}
\usepackage{mdwtab}
\usepackage{multirow}
\usepackage{multicol}
\usepackage{rotating}
\usepackage{setspace}
\usepackage{xspace}
\usepackage{amsmath}
\usepackage{algorithm}
\usepackage{algpseudocode}  
\begin{document}

\title{A Blockchain-based Platform Architecture for Multimedia Data Management}

\author{Yue Liu         \and
        Qinghua Lu \and
        \\Chunsheng Zhu \and
        Qiuyu Yu
}

\institute{Yue Liu \at
              School of Computer Science and Engineering, \\University of New South Wales, Australia\\
              Data61, CSIRO, Australia
            \and
           Qinghua Lu \at
             Data61, CSIRO, Australia
           \and
           Chunsheng Zhu \at
              SUSTech Institute of Future Networks,\\ Southern University of Science and Technology, China \\
              \email{chunsheng.tom.zhu@gmail.com}\\
              He is the corresponding author of this paper. 
           \and
           Qiuyu Yu \at
            College of Computer Science and Technology,\\ China University of Petroleum (East China), China
}

\date{Received: date / Accepted: date}

\maketitle

\begin{abstract}
Massive amounts of multimedia data (i.e., text, audio, video, graphics and animation) are being generated everyday. Conventionally, multimedia data are managed by the platforms maintained by multimedia service providers, which are generally designed using centralised architecture. However, such centralised architecture may lead to a single point of failure and disputes over royalties or other rights. It is hard to ensure the data integrity and track fulfilment of obligations listed on the copyright agreement. 
To tackle these issues, in this paper, we present a blockchain-based platform architecture for multimedia data management. We adopt self-sovereign identity for identity management and design a multi-level capability-based mechanism for access control. We implement a proof-of-concept prototype using the proposed approach and evaluate it using a use case. The results show that the proposed approach is feasible and has scalable performance.

\end{abstract}

\section{Introduction}
\label{intro}
Massive amounts of multimedia data (i.e., text, audio, video, graphics and animation) are being generated and used everyday. For example, Netflix users watch 164.8 million hours of video per day~\cite{netflix}. Copyright owners usually sign copyright agreements with multimedia service providers regarding the rights and royalties. How to ensure the fulfilment of obligations is an essential challenge which have tangled multimedia community for decades.

Conventionally, multimedia data are managed by the platforms maintained by multimedia service providers, which are generally designed using centralised architecture. However, such centralised architecture design may cause a single point of failure. Multimedia service providers may manipulate the data (e.g., number of downloads) without the copyright owners' knowledge, which can lead to disputes over royalties or other rights. One the other hand, multimedia service providers may be compromised, which consequently results in data tampering. Thus, it is hard to ensure the data integrity and track fulfilment of obligations listed on the copyright agreement. 

Blockchain is an innovative distributed ledger technology for building new forms of decentralized software architecture, which enables agreements on transactional data sharing across a large network of untrusted participants, without relying on a central trusted authority~\cite{scheuermann2015iacr}. The data structure of blockchain is a list of identifiable blocks, and all the blocks are linked to the previous block and thus formo a chain. The blocks are containers for storing transactions, which are identifiable packages carrying the changing states of data. A smart contract is a user-defined program, deployed and executed on a blockchain network~\cite{Omohundro:2014}. After deployment, a smart contract can express triggers, conditions to enable complex business logic.

Self-sovereign identity is an emerging identity management paradigm which enables entities to control the use of their data~\cite{pathToSSI}, which is aligned with the nature of blockchain. W3C has published a standard for decentralised identifiers (DIDs)~\cite{W3CDID} to support the concept of self-sovereign identity. DID contains human-readable information and can be used across different platforms. Specifically, a DID is a URL that refers to an entity for trusted interactions, and each DID bonds to a DID document (DDO), describing how to use that specific DID. 

In this paper, we adopt blockchain technology to build a platform for multimedia data management to ensure fulfilment of obligations in copyright agreement. The contributions of this paper are as follows. 
\begin{itemize}
 \item We present a blockchain-based platform architecture for multimedia data management to avoid the issue of single point of failure. All the signed copyright agreements are stored in the off-chain agreement repositories while the hash value of the agreements are stored in the agreement registry on-chain.
  
  \item We propose a self-sovereign identity method for identity management through DIDs, which could keep the copyright owners in the loop of data management. 
  
  \item We design a multi-level capability-based access control mechanism for multimedia data management. The access tokens are generated based on the signatures of both copyright owner and multimedia service providers as well as access period.
  
\end{itemize}
We implement the proposed approach in a proof-of-concept prototype and evaluate it using a use case. The evaluation results show that our proposed approach is feasible and has scalable performance.

The remainder of this paper is organised as follows. In the next section, we introduce some background knowledge and related works. Section 3 presents the overall architecture of blockchain-based platform for multimedia data management, and the design of provided services. Section 4 demonstrates the smart contract design in prototype implementation, and Section 5 evaluates the proposed solutions in terms of feasibility and performance. Section 6 concludes the paper and outlines the future work.

\section{Background and Related Works}
\label{Background}

\subsection{Blockchain}
Recently, blockchain has become a hot topic to develop next generation applications in a decentralised way, acting as both a distributed ledger and a computing infrastructure.

Being a distributed ledger in software architecture design, a blockchain network can verify and store transactions~\cite{scheuermann2015iacr}. Without relying on any central trusted authority, all participants in a blockchain network need to reach consensus on transactional data states to achieve trust. In the bitcoin consensus mechanism proposed by Nakamoto, the majority of nodes should be honest to against third party intermediaries through game theoretic incentives~\cite{Satoshi:bitcoin}. A blockchain is build up by a list of identifiable blocks linking to the previous one in a chronological order. Each block contains transactions which record the changing states of data. 

Smart contract enables on-chain computing, extending the capability of blockchain technology. Smart contracts~\cite{Omohundro:2014} are programs deployed and running on blockchain, which can express triggers, conditions and business logic to support more complex programmable transactions. Most blockchain platforms provide their own scripting language to develop smart contract, specially, Ethereum supports a Turing-complete language called Solidity\footnote{https://solidity.readthedocs.io/}.

Integrating blockchain technology into existing software architecture has been studied these years in both industry and academia. Some renowned companies provide blockchain services via cloud (e.g. Microsoft\footnote{\url{https://azure.microsoft.com/en-us/solutions/blockchain/}}, IBM\footnote{\url{https://www.ibm.com/blockchain/}}, Amazon\footnote{\url{https://aws.amazon.com/managed-blockchain}}). Blockchain has been considered as a storage and computation infrastructure and applied in multiple research fields, for instance, electronic health records~\cite{healthApp}, e-voting~\cite{eVoting}, industrial IoT~\cite{IIoT}, energy supply~\cite{energy}, etc. In previous works, we have designed a blockchain-based traceability application~\cite{IEEESoftware2017, originChain}, a unified blockchain as a service plarform~\cite{ubaas}, and platform architecture for blockchain-based self-sovereign identity~\cite{IEEESoftware2019}. In this study, we integrate blockchain into multimedia data management, providing a reference platform architecture to architects and developers.

\lstset{  
  frame=single,
  framesep=\fboxsep,
  framerule=\fboxrule,
  xleftmargin=\dimexpr\fboxsep+\fboxrule,
  xrightmargin=\dimexpr\fboxsep+\fboxrule,
  language=Java,
  basicstyle=\scriptsize\ttfamily,
  commentstyle=\color{cyan},
  tabsize=2,
  keywordstyle=,
  breaklines=true,  
  captionpos=b,
  escapeinside=``
}
\begin{lstlisting}[caption=DID document example.float,label=DDO]
{
  "@context": "https://w3id.org/did/v1",
  "id": "did:example:123456789abcdefghi",
  "publicKey":  "publicKey": [{
    "id": "did:example:123456789abcdefghi#keys-1",
    "type": "RsaVerificationKey2018",
    "controller": "did:example:123456789abcdefghi",
    "publicKeyPem": "-----BEGIN PUBLIC KEY...END PUBLIC KEY-----"
  }],
  "service": [{
    "id": "did:example:123456789abcdefghi#vcr",
    "type": "CredentialRepositoryService",
    "serviceEndpoint": "https://repository.example.com/service/8377464"
  }]
}
\end{lstlisting}

\subsection{Self-Sovereign Identity}
The era of big data has witnessed the increasing significance of digital identity management. As people have multiple relationships and act as different ``personas" in these relationships, identity management has become a fundamental requirement in our digitised society~\cite{modinis2005common,dabrowski2008generic}. However, Internet users do not generally have complete control over their digital identities stored by third-party issuers (e.g. social networking sites). The concept of self-sovereign identity~\cite{pathToSSI} is proposed to give entities the full administration of their identities (via personal mobile devices or cloud), and interact with others. The W3C Community Group specifies a guideline on the implementation of self-sovereign identity that each identity is presented by a decentralised identifier (DID)~\cite{W3CDID}, which contains human-readable information and can be used across different platforms. 

A DID can be considered as a URL which refers to the DID document (DDO), containing some basic information for trusted interactions. A simple DDO example given by W3C is presented in Listing~\ref{DDO}, which has four properties: \emph{context} denotes the DID specification between two DIDs; \emph{id} is the DID related to this DDO; \emph{publicKey} is for digital signatures and cryptographic operations; and \emph{service} list out the service endpoints that are used for interactions among DIDs.

DID and DDO do not contain detailed identity information. W3C proposes verifiable credentials which contains particular identity information attributes describing an entity. Within self-sovereign identity, entities can issue credentials with signature about themselves and others, which can then be shared and verified to identify themselves. Blockchain has been widely recognised as a viable technology to enable DID and verifiable credential. As self-sovereign identity eliminates the need of intermediary, it aligns with the decentralised nature of blockchain.

\subsection{Related Works}
There are already some researches considering blockchain as a solution to mitigate against the confinement of traditional centralised multimedia management scheme. Bhowmik and Feng~\cite{multimediaBlockchain} propose a blockchain-based watermarking framework, in which the watermark contains both transaction history and image hash value, for logging blockchian data states and ensuring data integrity respectively. Rathee et al.~\cite{multimediaIoT} focus on healthcare data processing in an IoT context, where blockchain is utilised to provide a transparent environment and trace historical data. Xu et al.~\cite{multimediaCloud} adopt blockchain in their cloudlet management method, securing the data integrity during the offloading procedure. Guo et al.~\cite{multimediaEducation} design a blockchain-based digital rights management for education multimedia resources, where a multi-chain architecture is applied across different educational networks. Vishwa and Hussain~\cite{multimediaPrivacy} propose a blockchain-based protocol protect and multimedia privacy, while Lee and Park~\cite{multimediaMerkle} use blockchain to ensure the integrity and security of video data. Compared with the related works, this study presents a more general approach for the architecture design of blockchain-based multimedia management, which covers the registration and access control of multimedia resources. The concept of self-sovereign identity is integrated for identity management. Furthermore, a multi-level access control mechanism is demonstrated to provide a detailed overview of user interactions.

\section{Blockchain-based Multimedia Data Management}
\label{Architecture}

In this section, we present our approach for multimedia data management using blockchain technology. First, the multimedia management process is analysed. Secondly, the overall architecture is presented to give a systematic view of blockchain-based multimedia data management design. Thirdly, the identity management and access control mechanisms are proposed.

\begin{figure*}[t]
\begin{center}
\includegraphics[width=\textwidth]{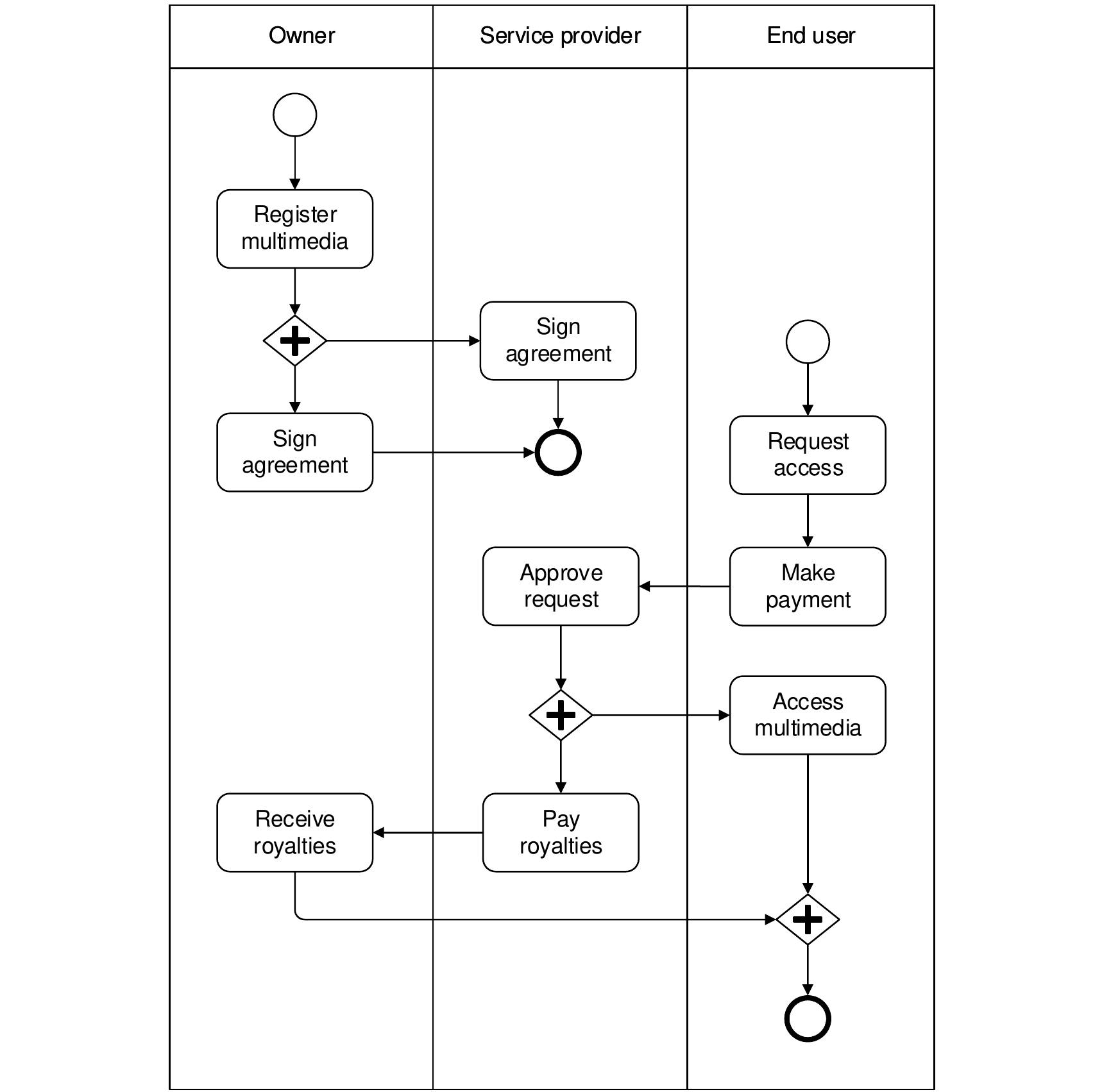}
\caption{Multimedia Management Process.} 
\label{lifecycle-pic}
\end{center}
\end{figure*}

\begin{figure*}[t]
\begin{center}
\includegraphics[width=0.9\textwidth]{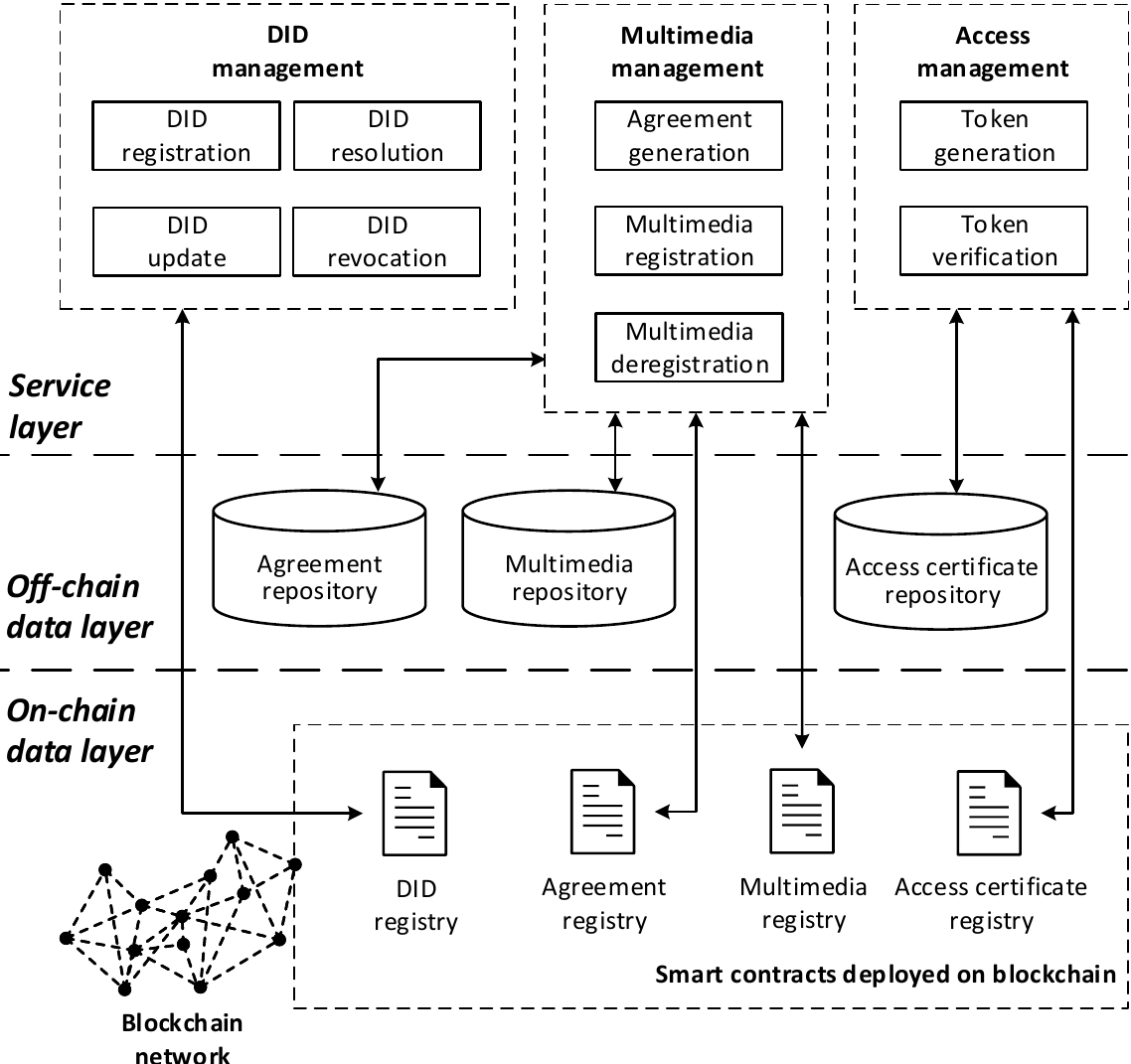}
\caption{Architecture for Blockchain-based Multimedia Data Management.} 
\label{architecture-pic}
\end{center}
\end{figure*}

\subsection{Multimedia Management Process}
Multimedia works consist of different data types, including diagrams, text, videos, audios, etc., Fig.~\ref{lifecycle-pic} illustrates the multimedia management process, in which each task indicates the services that should be supported in a multimedia data management platform.

First, a multimedia work is created and the creation team or company then becomes the owner of the multimedia work. The multimedia work (i.e. data) can be registered in a multimedia management platform to protect the owner's rights when spreading the inner idea or making profit. After the inspection of content by multimedia service provider, the two involving parties should sign an agreement on the copyright licensing agreement. 

A registered multimedia may be requested by end users for different purposes (e.g. reproduction, performance, broadcasting, etc.), which needs the authorisation of service provider, which acts as an agent for multimedia owner. The end user needs to make payment when requiring the access rights, and it can access the source file and other related document of the multimedia after obtaining service provider's permission. Meanwhile, the service provider should pay royalties to the multimedia owner according to their agreement (e.g. the number of access).

\subsection{Architecture Design}
Fig.~\ref{architecture-pic} shows the overall architecture design from the perspective of platform users, including multimedia owner, service provider, and end user, which follows the analysis of management process above. The architecture consists of three layers: service layer, off-chain data layer, and on-chain data layer. First, The service layer is comprised of different management service: \textit{DID management}, \textit{Multimedia management}, and \textit{Access management}, and each type of service is provided by at least one smart contract, to enable different functionalities. Every platform user should have at least one DID to identify itself via \textit{DID management}, and each DID is bound with a blockchain account to preserve uniqueness. The multimedia service agreement and data registration are realised by \textit{Multimedia management}, while \textit{Access management} grants and verifies specific access rights. Secondly, off-chain data layer stores raw multimedia files and documents, and also access certificates which contain authorisation information of rights to particular multimedia. Finally, a blockchain network and smart contracts are applied in the architecture design, to process on-chain data and business logic.


\subsection{Protocols for Multimedia Data Management}
This section describes key protocols designed for multimedia data management, including DID management process, copyright licensing process, access granting process, and version control process. Please note that in this study, we focus on the on-chain business logic as the traditional multimedia data management scheme are mature.
 
\subsubsection{DID Management for Self-Sovereign Identity}
Each user needs to register a DID using a blockchain account through \textit{DID registration}. A registered DID represents the identity of a user within the platform. Note that a blockchain transaction is created when invoking this service, to send DDO to the blockchain. The DID is recorded with its corresponding DDO in \textit{DID registry} after successful registration.

Registered DID can be viewed by \textit{DID resolution}, which illustrates the DDO to users. The DDO contains a user's public key to verify its digital signature and service endpoints for personal interaction. If the DDO data is changed, the user can invoke \textit{DDO update} to upload new information. Finally, a DID can be revoked when it has no use, and the stored DDO is then deleted from \textit{DID registry}. Please note that \textit{DDO update} and \textit{DID revocation} can only be carried out by the DID owner who has the control of related blockchain account, any other malicious operation to falsify a DID is denied.

\begin{figure*}[t]
\begin{center}
\includegraphics[width=0.95\textwidth]{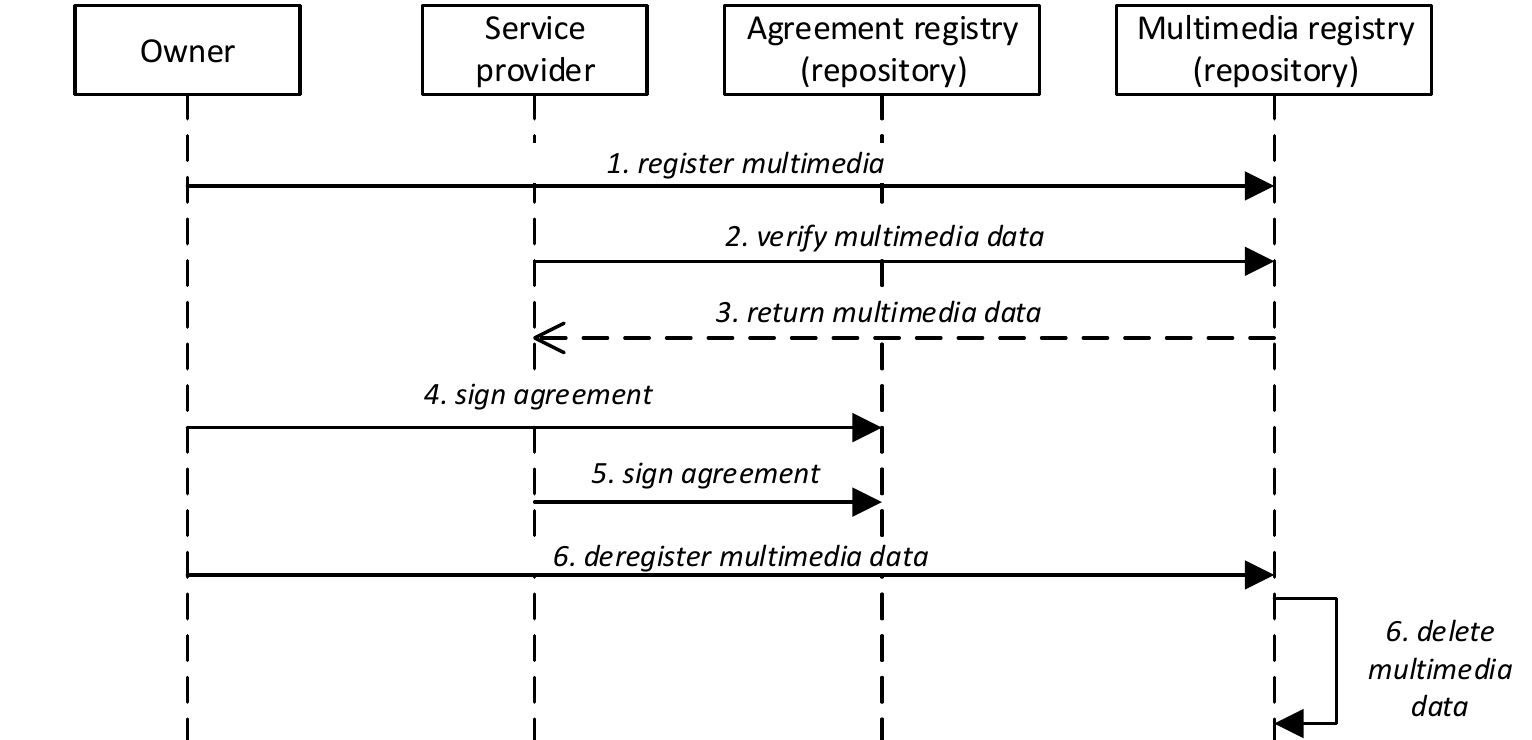}
\caption{Copyright licensing process.} 
\label{multi-pic}
\end{center}
\end{figure*}

\subsubsection{Copyright Licensing}
Fig.~\ref{multi-pic} illustrates the user interactions over multimedia data management. A multimedia owner uploads the sensitive information of multimedia data to \textit{Multimedia registry} via \textit{Multimedia registration}, while the original file is stored in off-chain \textit{Multimedia repository}, as the current blockchain technology does not support the storage of multimedia data like figures or videos. The registration needs to be inspected by service provider, which can send employees to manually verify whether the multimedia meets requirements, or apply its existing automatic procedures. The next step is that the multimedia owner and service provider discuss the business details and sign an agreement, which is then recorded in \textit{Agreement registry}. Moreover, the owner can decide to delete the multimedia by \textit{Multimedia deregistration}.

\begin{figure*}[t]
\begin{center}
\includegraphics[width=\textwidth]{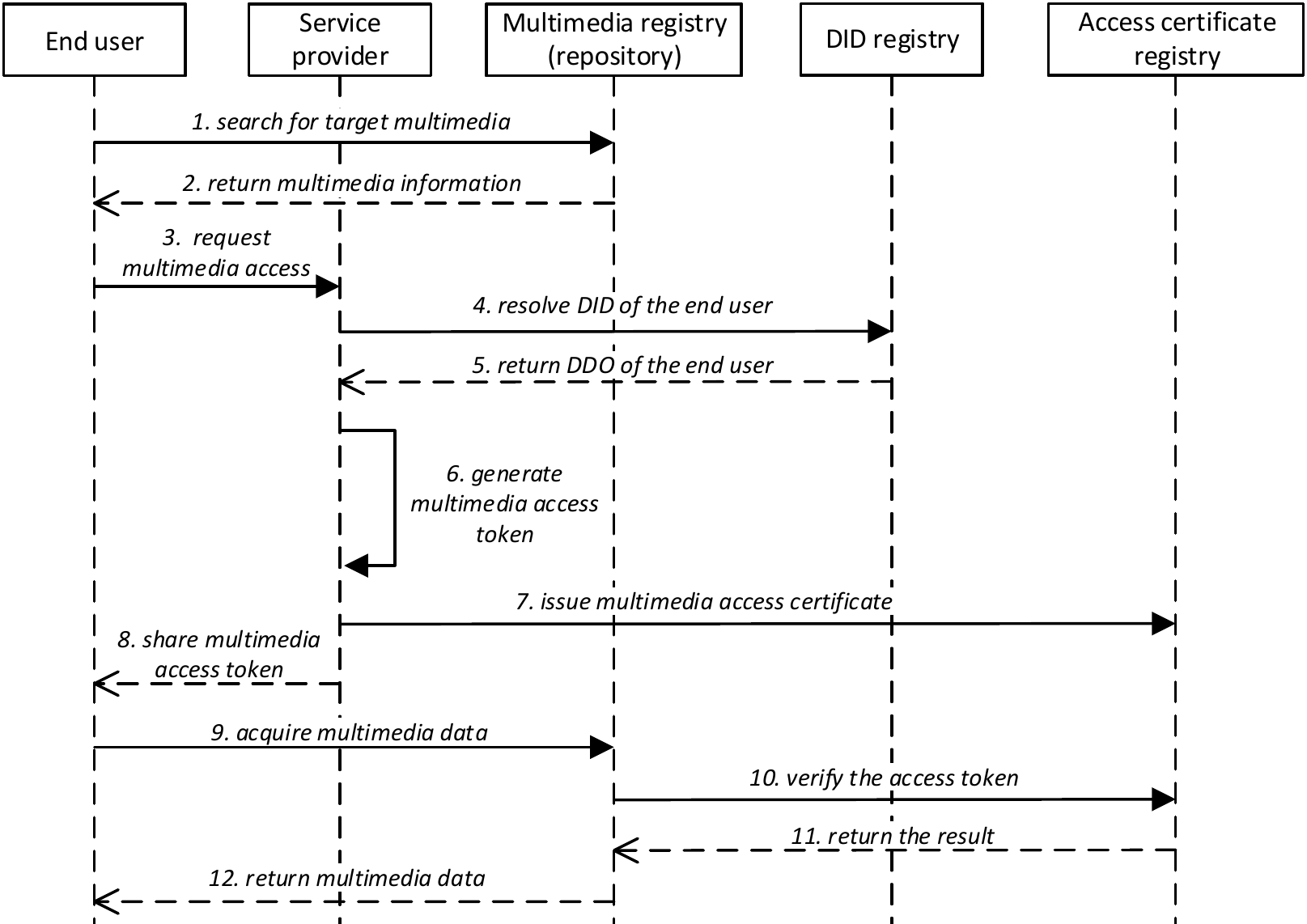}
\caption{Access granting process.} 
\label{share-pic}
\end{center}
\end{figure*}

\subsubsection{Access Granting}
\label{granting}

The access granting process of a multimedia data is shown in Fig.~\ref{share-pic}. An end user first searches for target multimedia, and send requests for specif access rights from service provider. The service provider obtains DDO information of this end user for the generation of access token, and issues corresponding certificate to \textit{Access certificate registry}. 


\begin{equation}
\begin{split}
token_{\emph{access}} \rightarrow \left\{
\begin{aligned}
&\textit{Owner}, \textit{Provider}, \textit{Enduser}, \textit{Multimedia}, \textit{Access Rights},\\
&  \textit{Valid Time}, \textit{OwnerSignature}, \textit{ProviderSignature}
\end{aligned}
\right.
\end{split}
\end{equation}

\begin{itemize}
  \item[·] \textit{Owner}: the DID of multimedia owner; 
  
   \item[·] \textit{Provider}: the DID of multimedia service provider, who is granting the permission to \textit{Enduser} to conduct further operations on \textit{Multimedia};
  
  \item[·] \textit{Enduser}: the DID of multimedia end user, who is applying for the access control of multimedia data;
  
  \item[·] \textit{Multimedia}: the ID of multimedia data, possessed by the \textit{Owner}; 
  
  \item[·] \textit{Access Rights}: specific access rights to the \textit{Multimedia};
  
  \item[·] \textit{Valid Time}: the time period within which the access rights are valid, once expired, the \textit{Requestor} will not be allowed to use the \textit{Multimedia};
  
  \item[·] \textit{OwnerSignature}: the \textit{Owner} signature of chosen \textit{Multimedia};
  
  \item[·] \textit{ProviderSignature}: the \textit{Provider}'s signature over the above attributes.
\end{itemize}

\begin{algorithm}[h!]
\caption{Generating and Verifying Token}  
  \label{generateverifytoken}
  \footnotesize
  \begin{algorithmic}[1]
  \State /*Generating Token*/
  \State $Own, OwnSig \gets$ search($MulID$)
  \State $OnchainInfo \gets$ combine($Own, Pro, User, MulID, ARs, Time, OwnSig$)
  \State $CertID \gets$ hash($OnchainInfo$)
  \State $ProSig \gets$ sign($OnchainInfo$)
  \State $token \gets$ generate($CertID, Pro, ProSig, Time$)
  \State /*To access certificate registry*/
  \State issue($OnchainInfo, ProSig$)
  \\
  \State /*Verifying Token*/
  \State $CertID, Pro, ProSig, Time \gets$ decode($token$)
  \If{expired($Time$) == true}
    \State /*$Pro, User, ARs, Time, OwnSig, ProSig$ are on-chian information*/
    \State $Pro, User, ARs, Time, OwnSig, ProSig \gets$ search($CertID$)
    \If{equal($ProSig$(on-chian), $ProSig$) == true}
      \State $ProPubKey \gets$ resolve($Pro$)
      \State $OnchainInfo \gets$ combine($Own, Pro, User, MulID, ARs, Time, OwnSig$)
      \State verify($ProPubKey, OnchainInfo, ProSig$)
    \EndIf
  \EndIf
  \end{algorithmic}
\end{algorithm}

After the issuance of access certificate, service provider shares the multimedia access token to end user via the contact information stored in DDO. When accessing the multimedia data, the end user needs to present its token for verification, which involves three steps: ensure the token is not expired, verify the provider's signature, and verify the owner's signature. Only authenticated end users are capable to obtain multimedia data (e.g. source file).

Algorithm~\ref{generateverifytoken} illustrates a more detailed process of token generation and verification. The service provider obtains the owner DID and signature through the ID of requested multimedia, and the hash of owner DID, provider DID, end user DID, multimedia ID, granting access rights, valid time period, owner signature is combined to generate the certificate ID. After signing the above information, the token is created using certificate ID, provider DID, provider signature, and valid time period. When a token is being verified, it is first decoded to extract inner information, and the verification is failed if the token is expired. The decoded provider signature is compared with the on-chain provider signature to ensure integrity, after which, the signature is verified with provider's public key acquired in its DDO.

\section{Prototype Implementation}
\label{implementation}
The implementation details are introduced in this section. The approach prototype is developed using Node.js\footnote{\url{https://nodejs.org/}} version 10, and all involved participants can send requests to acquire specific services via RESTful APIs which are implemented using express.js\footnote{\url{http://expressjs.com/}} server. In this study, we apply Parity\footnote{\url{https://www.parity.io/}} 1.9.2 consortium blockchain as the underlying blockchain network, and the smart contracts are scripted using Solidity language.

\begin{lstlisting}[caption=DID registry implementation.,label=DIDCode]
contract DIDRegistry{
    mapping(string => bool) registered;
    struct DDO{
        address owner;
        address id;
        string did;
        string ddo;
    }
    mapping(string => DDO) private identity;
    function register(address _id, string _did, string _ddo) public{
        require(registered[_did] == false);
        registered[_did] = true;
        identity[_did].owner = msg.sender;
        identity[_did].id = _id;
        identity[_did].did = _did;
        identity[_did].ddo = _ddo;
    }
    modifier onlyOwner(string _did){
        require(msg.sender == identity[_did].owner);
        _;
    }
    function updateDDO(string _did, string _ddo) public onlyOwner(_did){
        identity[_did].ddo = _ddo;
    }
    function resolve(string _did) public view returns (string __ddo){
        return (identity[_did].ddo);
    }
    function revoke(string _did) public onlyOwner(_did){
        delete identity[_did];
        registered[_did] = true;
    }
}
\end{lstlisting}

Listing~\ref{DIDCode} presents the design of \textit{DID registry} smart contract, which adheres Ethereum Request for Comments (ERC) 1056 standard\footnote{{\url{https://github.com/ethereum/EIPs/issues/1056}}}. ERC standards are proposed by Ethereum contributors to promote the development of Ethereum-based applications, while ERC1056 describes a lightweight registry to manage blockchain-based digital identities. In \textit{DID registry}, all registered DID and DDO are recorded, which follows the W3C standard specification~\cite{W3CDID}. Each service is mapped to a function, while ``updateDDO" and ``revoke" functions are both restricted by the modifier ``onlyOwner", which means that these two functions can only be called by the DID owner who controls the correct blockchain account.

\begin{lstlisting}[caption=Agreement registry implementation.,label=AgreementCode]
contract AgreementRegistry{
    mapping(string => bool) signed;
    struct agreement{
        string id;
        string owner_did;
        ...
    }
    mapping(string => agreement) private agreement_info;
    function generate(string _id, string _owner_did, address _owner, string _provider_did, string _agreementHash, uint _valid_time, string _copyrights) public{
        require(signed[_id] == false);
        agreement_info[_id].id = _id;
        agreement_info[_id].owner_did = _owner_did;
        ...
    }
    modifier onlyOwner(string _id){
        require(msg.sender == agreement_info[_id].owner);
        _;
    }
    modifier onlyProvider(string _id){...}
    function ownerSign(string _id, string _sig) public onlyOwner(_id){
        require(signed[_id] == false);
        require(agreement_info[_id].owner_signed  == false);
        agreement_info[_id].owner_sig = _sig;
        agreement_info[_id].owner_signed = true;
        if(agreement_info[_id].provider_signed == true)
          signed[_id] = true;
    }
    function providerSign(string _id, string _sig) public onlyProvider(_id){...}
}
\end{lstlisting}

Listing~\ref{AgreementCode} shows the script of \textit{Agreement registry}. When multimedia owner and service provider sign the legal agreement, the related information need to be stored on-chain to ensure integrity, including the identity of involving two parties, the hash value of agreement file, and the signatures of both parties on the agreement. When the two signatures are recorded, the on-chain agreement is settled and not allowed for revision.

Listing~\ref{MultimediaCode} demonstrates the code of \textit{Multimedia registry} smart contract, which contains a struct to record the multimedia information. After the registration application of a multimedia, the owner and service provider sign legal agreement about their cooperation, and provider approves this registration by uploading the hash value of legal agreement. This contract also logs the access control information of each multimedia resource. A registered multimedia can be deregistered by its owner, and the on-chain information is deleted upon this operation.

Listing~\ref{accessCode} is the implementation of \textit{Access certificate registry}. When an end user requests for particular access rights of a multimedia, the service provider should issue a certificate to this smart contract. On-chain access control certificate data is mapped to off-chain access token generated by multimedia owner. The token is implemented in the form of JSON Web Token (JWT), which can be used to access and verify a certificate for the access rights of an end user when acquiring multimedia resources. A generated token is also recorded in \textit{Multimedia registry} for future auditing work.

\begin{lstlisting}[caption=Multimedia registry implementation.,label=MultimediaCode]
contract MultimediaRegistry{
    mapping(string => bool) registered;
    struct multimedia{
        string id;
        string owner_did;
        address owner;
        ...
    }
    mapping(string => multimedia) private multimedia_info;
    function register(string _id, string _owner_did, string _hash, string _sig, string _upload) public{
        require(registered[_id] == false);
        registered[_id] = true;
        multimedia_info[_id].id = _id;
        multimedia_info[_id].owner_did = _owner_did;
        ...
        }
    }
    modifier onlyOwner(string _id){
        require(msg.sender == multimedia_info[_id].owner);
        _;
    }
    function deregister(string _id) public onlyOwner(_id){
        delete multimedia_info[_id];
    }
    modifier onlyProvider(string _id){
        require(msg.sender == multimedia_info[_id].provider);
        _;
    }
    function approve(string _id, string _did, string _agreementHash) public onlyProvider(_id){
        require(registered[_id] == true);
        require(multimedia_info[_id].approved == false);
        multimedia_info[_id].approved = true;
        multimedia_info[_id].provider = msg.sender;
        ...
    }
    function accessInfo(string _id, string _certID) public onlyProvider(_id){
         multimedia_info[_id].access_info.push(_certID);
    }
    ...
}
\end{lstlisting}


\section{Evaluation}
\label{Evaluation}

In this section, we evaluate the proposed approach in terms of feasibility and performance. First, we describe the use case of our approach, data rights management for digital music resources. Afterwards, the feasibility of applying the proposed design in the use case is evaluated, and we discuss how this design can improve the security of whole platform. Finally, the performance of core services are tested with quantitative experiments.

\begin{lstlisting}[caption=Access certificate registry implementation.,label=accessCode]
contract accessCertificateRegistry{
    struct multimediaAccess{
        string id;
        string provider_did;
        string enduser_did;
        ...
    }
    mapping(string => multimediaAccess) private certificate;
    function issueCert(string _id, string _multimedia_id, string _provider_did, string _enduser_did, string _owner_sig, uint access_right, uint _valid_time, string _sig) public{
        require(certificate[_id].shared == false);
        certificate[_id].id = _id;
        certificate[_id].multimedia_id = _multimedia_id;
        certificate[_id].provider_did = _provider_did;
        certificate[_id].enduser_did = _enduser_did;
        ...
    }
    ...
}
\end{lstlisting}

\subsection{Use Case}
In the feasibility evaluation, we select data rights management (DRM) for digital music resources as the use case. Generally, the data rights of digital music resources include the right of publication, revision, reproduction, exhibition, performance, etc., and owners can authorise other people on these rights for different purposes~\cite{copyrightLaw}.

In existing music DRM platforms, the authors of music upload their works to a platform, which issues a legal certificate to the authors after successful registration of music works. Other artists can apply for specific copyrights from the platform, and wait for the permission.

By applying the proposed architecture design, a music DRM platform can still preserve most of the current workflow, but needs to transfer some workflow and data storage to on-chain smart contracts. A music author should first register DID to identify itself with its blockchain account. When registering an original music work to platform, the raw music files are stored in off-chain \textit{Multimedia repository}, while the basic information is sent to \textit{Multimedia registry}. The platform users (audiences) should also have their own DID when using the platform, and propose requests of particular copyrights for different purposes. If platform agrees to authorise required copyrights, it needs to share the access token, and issue an on-chain certificate, which contains detailed information of the involved three parties, music work, and copyrights, to \textit{Access certificate registry}. End users can access the music or other raw files after the verification of received token. 


\subsection{Feasibility}
The feasibility of our architecture design is manifested via the use case of music DRM platform. Integrating blockchain technology into the data rights sharing of music works can improve the security of whole platform, in terms of confidentiality, integrity, and availability~\cite{CIATriad}. First, platform users are identified by DID and related blockchain account instead of traditional personal ID and password instead of personal ID and password stored in traditional database. Each blockchain account has local files in the personal device, which ensures the confidentiality of data. Meanwhile, on-chain smart contracts have coded mechanism, which only allows authorised users to conduct specific operations (e.g. only DID owner can update and revoke DID). The access rights to a music work is recorded on blockchain, and needs to be verified before multimedia acquisition. Secondly, in the architecture design, we apply an on-chain \& off-chain pattern to store the raw files and documents off-chain, while sensitive and important information (e.g. hash value of the music source file) is sent on-chain. Such separation of storage ensures data integrity that one can hash the file after acquisition, and compare it with on-chain hash value. Finally, blockchain contains the log of user operations related to DID, which can be used for the auditing works. Moreover, each blockchain network participant keeps a local replica of all historical transactions. Consequently, the platform can be recovered using the replicas even a node is attacked and out-of-service.

\subsection{Performance}
To evaluate the performance of proposed approach, we conducted experiments to measure the response time of seven core services in the architecture design: \textit{DID registration}, \textit{DID resolution}, \textit{Agreement generation}, \textit{Multimedia registration}, \textit{Multimedia deregistration}, \textit{Token generation}, and \textit{Token verification}. 

We deployed the proof-of-concept prototype on Alibaba Cloud\footnote{\url{https://www.aliyun.com/}} virtual machine (VM) (4 vCPUs, 16G RAM, 40GB disk), and adopted Parity consortium blockchain 1.9.3-stable, in which the consensus algorithm is Proof-of-Authority (PoA). The block gas limit is set to 80M and block interval is configured to 5s. The smart contracts are written in Solidity with compiler v.0.4.26. We performed seven tests to measure the response time of aforementioned functions respectively. The API requests, containing required information of these services, are produced via JMeter running on another VM as load generator (8 vCPUs, 8G RAM, 40GB disk). In each test, we send 10,000 requests in total, and JMeter is configured to 20 creations per batch (API calls).

\begin{figure*}[t]
\begin{center}
\includegraphics[width=\textwidth]{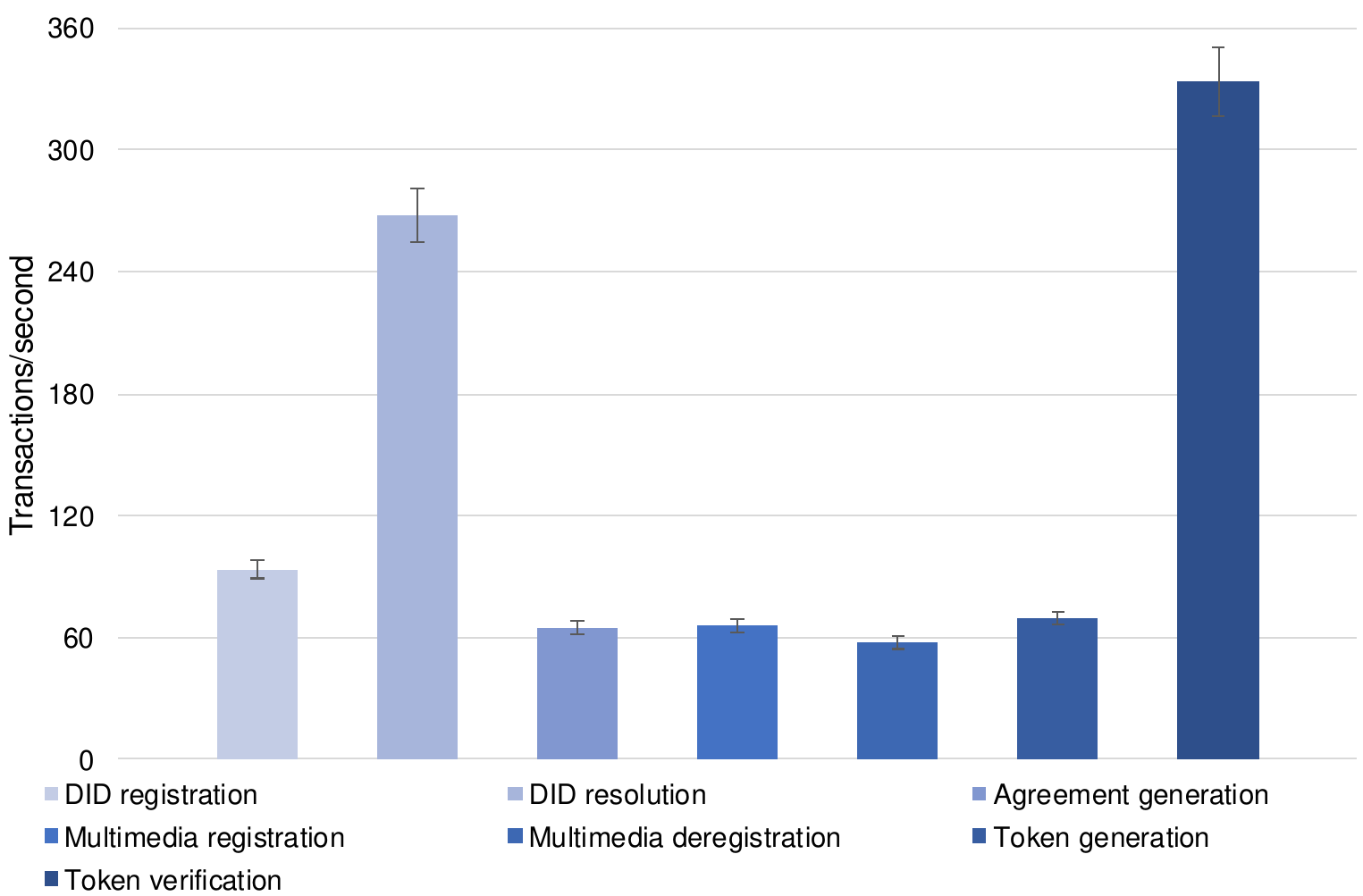}
\caption{Performance evaluation.} 
\label{data}
\end{center}
\end{figure*}

Fig.~\ref{data} shows the service throughput in terms of the number of many requests completed per second. The x-axis represents different services measured in the tests, while the y-axis represents the average throughput (transaction per second - tps) over 10,000 API requests. \textit{Agreement generation}, \textit{Multimedia registration}, \textit{Multimedia deregistration}, and \textit{Token generation} reach a throughput range of 57-70 tps, while the average throughput of \textit{DID registration} is about 93 tps. \textit{DID resolution} and \textit{Token verification} have higher throughput than the other services, which are around 268 tps and 333 tps respectively.

Compared the highest two services, \textit{DID registration} \textit{Agreement generation}, \textit{Multimedia registration}, \textit{Multimedia deregistration}, and \textit{Token generation} all change the data states of blockchain (storing or deleting data). Such operations generate on-chain transactions that need to be included in a block, which is then affected by the block interval predefined during blockchain deployment. Further, as the capability of a block is settled, it includes different number of transactions for these five services, which results in the difference of throughput. \textit{DID resolution} and \textit{Token verification} only obtain required information from smart contracts, and do not generate blockchain transactions. Hence, they are less time-consuming when processing API requests. From the experiment results, it can be observed that our platform achieves scalable performance.

\section{Conclusion}
\label{Conclusion}
This paper presented a platform architecture for blockchain-based multimedia data management, to tackle with the single point failure and infringement problems for the centralised design. We first discussed the multimedia management process to determine the core services which should be included in the architecture, and then presented a three-layer architecture which consists of service layer, off-chain data layer, and on-chain data layer. Afterwards, a capability-based access control mechanism for multimedia data is proposed, where the concept of self-sovereign identity is adopted for identity management. The mechanism provides a guide of user interactions according to the multimedia management process. The proposed solution is evaluated via a use case, which illustrates that the proposed approach is feasible. Moreover, quantitative experiments are conducted, and the results show that the solution has scalable performance.

In the future, we plan to design the off-chain data layer using IPFS\footnote{\url{https://ipfs.io/}}, to further enhance the decentralisation of whole architecture. Further, we will extend the design with payment support. 



%
%




%
%

\end{document}